# Comment on "Modeling oxygen self-diffusion in UO$_2$ under pressure by M.W.D Cooper et al., Solid State Ionics 282 (2015) 26-30"

N. V. Sarlis and E. S. Skordas*

Department of Solid State Physics and Solid Earth Physics Institute, Faculty of Physics, School of Science, National and Kapodistrian University of Athens, Panepistimiopolis, Zografos 157 84, Athens, Greece.

**Abstract**

The oxygen self-diffusion coefficient in UO$_2$ has been recently studied [Cooper et al. Solid State Ionics 282 (2015) 26-30] over a range of pressures (0-10GPa) and temperatures (300-1900K) by combining molecular dynamics calculations with a thermodynamical model, the cBΩ model. A significant reduction in oxygen self-diffusion as a function of increasing hydrostatic pressure, and the associated increase in activation energy was identified. Here, we extend this study and find that the compressibility of the corresponding activation volume exceeds significantly the compressibility of the bulk material by almost one order of magnitude. This results is important since in the literature it is usually assumed that these two compressibilities are equal. The same holds when comparing the thermal expansion coefficient of this volume with that of the bulk solid.

## 1. Introduction

In order to reduce greenhouse gas emissions, an increase of the use of nuclear energy is recently reconsidered. The main component of conventional nuclear fuel is $UO_2$, which can be also mixed with other actinide oxides, for example $ThO_2$ and $PuO_2$, to form mixed oxide fuel [1-3]. Very recently the oxygen self-diffusion coefficient has been studied [4-6] in these materials that are important to nuclear fuel applications. In particular, the temperature variation of this coefficient in $UO_2$ and $ThO_2$ has been investigated in the temperature range 2000 to 3000 K [4]. As for the pressure variation of oxygen self-diffusion in $UO_2$ over a range of pressures 0-10 GPa, it just appeared by Cooper et al [6] who combined molecular dynamics (MD) calculations with a thermodynamical model, termed $cB\Omega$ model (see below). Specifically, MD calculations were carried out using the large-scale Atomic/Molecular Massively Parallel Simulator [7] by employing the CRG potentials [8] that have been found to reproduce [9] the thermomechanical and thermophysical properties of $AmO_2$, $CeO_2$, $CmO_2$, $NpO_2$, $PuO_2$, $ThO_2$ and $UO_2$ and be of usefulness in calculating the diffusion properties in $CeO_2$, $U_{1-x}Th_xO_2$ and $Pu_{1-x}U_xO_2$ [8, 9, 10].

According to the so called $cB\Omega$ model [11, 12] the defect Gibbs energy $g^i$ is interrelated with the bulk properties of the solid through the relation:

$$g^i = c^i B \Omega \qquad (1)$$

where $c^i$ stands for a dimensionless constant, $B$ is the isothermal bulk modulus and $\Omega$ the mean volume per atom. The superscript $i$ refers to the defect process [13] under investigation, e.g. defect formation, defect migration, self-diffusion activation. This model has been successfully applied for various defect processes (for a review see Ref. 12) to several categories of solids including metals [11], fluorides [14], diamond [15], mixed alkali halides [16, 17], semiconductors [18] as well as to materials that under uniaxial stress emit electric signals before failure [19] in a similar fashion to those detected before earthquakes [20, 21].

In the study of Cooper et al [6] both the expansivity and $B$ were derived using MD for the target range of temperatures, $T$=300 - 1900 K, and pressures $P$=0 - 10 GPa. They found the following expression for $B(T,P)$ that matches the full set of MD data they computed:

$$B(T,P) = a + bT + cT^2 + dP + eP^2 + fPT \qquad (2)$$

where $a$=218.0 GPa is the bulk modulus at $T$=0, $P$=0, $b$=-4.33×10$^{-2}$ GPa K$^{-1}$, $c$=-1.846×10$^{-6}$ GPa K$^{-2}$, $d$=5.864 is the pressure derivative of $B$, i.e., $\frac{dB}{dP}$, at $T$=0, $P$=0, $e$=-1.387×10$^{-1}$ GPa$^{-1}$ and $f$=1.301×10$^{-3}$ K$^{-1}$ (see Table 1 of Cooper et al [6])

Using the above MD data for $B(T,P)$ and adopting -according to Eq.(1)- for the self-diffusion process the expression

$$g^{act} = c^{act} B \Omega \qquad (3)$$

for the Gibbs activation energy, Cooper et al. [6] found that the following relation associates the oxygen diffusion coefficients to the isothermal bulk modulus and the mean volume $\Omega$ per atom in UO$_2$:

$$D^{UO_2}_{cB\Omega} = 1.277 e^{-\frac{0.3052 B \Omega}{k_B T}} 10^{-4} m^2 s^{-1} \qquad (4)$$

where $k_B$ denotes the usual Boltzmann constant. As for the activation volume

$$\upsilon^{act} = \left. \frac{\partial g^{act}}{\partial P} \right|_T \qquad (5)$$

they found that $\upsilon^{act} = (8.75 - 10.66) \times 10^{-6} m^3 mol^{-1}$ over the temperature range 700-1500 K. It is the scope of the present short paper to quantify the pressure and temperature dependence of $\upsilon^{act}$.

## 2. Additional comments deduced from the combination of the cBΩ model with MD.

The compressibility $\kappa^{act}$ and the thermal expansion coefficient $\beta^{act}$ of the activation volume $\upsilon^{act}$ are defined as follows:

$$\kappa^{act} = -\frac{1}{\upsilon^{act}}\frac{d\upsilon^{act}}{dP}\bigg|_T \quad (6)$$

$$\text{and } \beta^{act} = \frac{1}{\upsilon_{act}}\frac{d\upsilon^{act}}{dT}\bigg|_P \quad (7)$$

By inserting Eq.(3) into Eq.(5) and then using Eqs.(6) and (7), respectively we find [12]

$$\kappa^{act} = \kappa - \frac{\frac{d^2B}{dP^2}\big|_T}{\frac{dB}{dP}\big|_T - 1} \quad (8)$$

$$\text{and } \beta^{act} = \beta + \frac{\frac{d}{dT}\left\{\frac{\partial B}{\partial P}\big|_T\right\}}{\frac{dB}{dP}\big|_T - 1} \quad (9)$$

where $\kappa$ and $\beta$ stand for the isothermal compressibility and the volume thermal expansion coefficient of the bulk solid defined as $\kappa = -\frac{1}{V}\frac{dV}{dP}\big|_T$ and $\beta = \frac{1}{V}\frac{dV}{dT}\big|_P$, respectively. Equations (8) and (9), which enable the calculation of $\kappa^{act}$ and $\beta^{act}$ in terms of the bulk elastic and expansivity data, can be rewritten as:

$$\frac{\kappa^{act}}{\kappa} = 1 - \frac{B\frac{d^2B}{dP^2}\big|_T}{\frac{dB}{dP}\big|_T - 1} \quad (10)$$

$$\text{and } \frac{\beta^{act}}{\beta} = 1 + \frac{\frac{d}{dT}\left\{\frac{\partial B}{\partial P}\big|_T\right\}}{\beta\left(\frac{dB}{dP}\big|_T - 1\right)}, \quad (11)$$

respectively. By considering Eq.(2), the use of Eq.(10) leads to the calculation of the ratio $\frac{\kappa^{act}}{\kappa}$ at various temperatures. Indicative results for several pressures are given in Fig. 1. An inspection of this figure reveals that $\frac{\kappa^{act}}{\kappa}$ exceeds unity, as expected from Eq.(10) because in its last term we have [12]: $\frac{d^2B}{dP^2}<0$ in the numerator and $\left.\frac{dB}{dP}\right|_T >1$ in the denominator. Moreover, $\kappa^{act}$ exceeds $\kappa$ significantly, i.e., almost one order of magnitude or so. This is important because it is usually assumed in the literature [22] that the activation volume changes with pressure in the same manner as crystal volume, i.e., $\kappa^{act}=\kappa$ [23].

In addition, by applying Eq.(11) we calculate the ratio $\frac{\beta^{act}}{\beta}$ at various temperatures and pressures. Indicative results are given in Fig. 2, where we plot $\frac{\beta^{act}}{\beta}$ versus temperature for constant pressure. Results are depicted for several values of the pressure shown with different colours (see the inset). In this application, we considered that $\beta$ can be calculated from a relation:

$$\Omega = \Omega_0 + a_1 P + a_2 P^2 + \beta_1 T + \beta_2 T^2 + \gamma PT \quad (12)$$

similar to that of Eq.(2), which results from a polynomial fit to the data presented in Fig.1 of Cooper et al. [6] yielding $\Omega_0 = 13.513 \text{Å}^3$, $\alpha_1 = -6.387 \times 10^{-2} \text{Å}^3 \, GPa^{-1}$, $\alpha_2 = 1.267 \times 10^{-3} \text{Å}^3 \, GPa^{-2}$, $\beta_1 = 3.881 \times 10^{-4} \text{Å}^3 \, K^{-1}$, $\beta_2 = 4.6 \times 10^{-8} \text{Å}^3 \, K^{-2}$, and $\gamma = -1.662 \times 10^{-5} \text{Å}^3 \, GPa^{-1}K^{-1}$ and gives bulk moduli compatible within ±5% to those obtained from Eq.(2) for P<6GPa and T≤1360K. An inspection of Fig. 2 shows that the ratio $\frac{\beta^{act}}{\beta}$ exceeds unity as expected from Eq. (11) when considering that the numerator of the last term is positive since the value of $\left.\frac{\partial B}{\partial P}\right|_T$ increases upon increasing the temperature, which stems from anharmonic effects [12]. In addition, $\beta^{act}$ is almost one order of magnitude than $\beta$.

## 3. Conclusions

Here, we extended the results of the study of Cooper et al [6] that have been deduced from a combination of MD calculations with the cBΩ model. In particular, we determined the compressibility of the oxygen self-diffusion activation volume and found that it significantly exceeds the compressibility of the bulk material. This conclusion is important since it is usually assumed in the literature that these two quantities are equal. In addition, we find that the thermal expansion coefficient of the activation volume is significantly larger than the corresponding coefficient of the bulk volume.

Figures

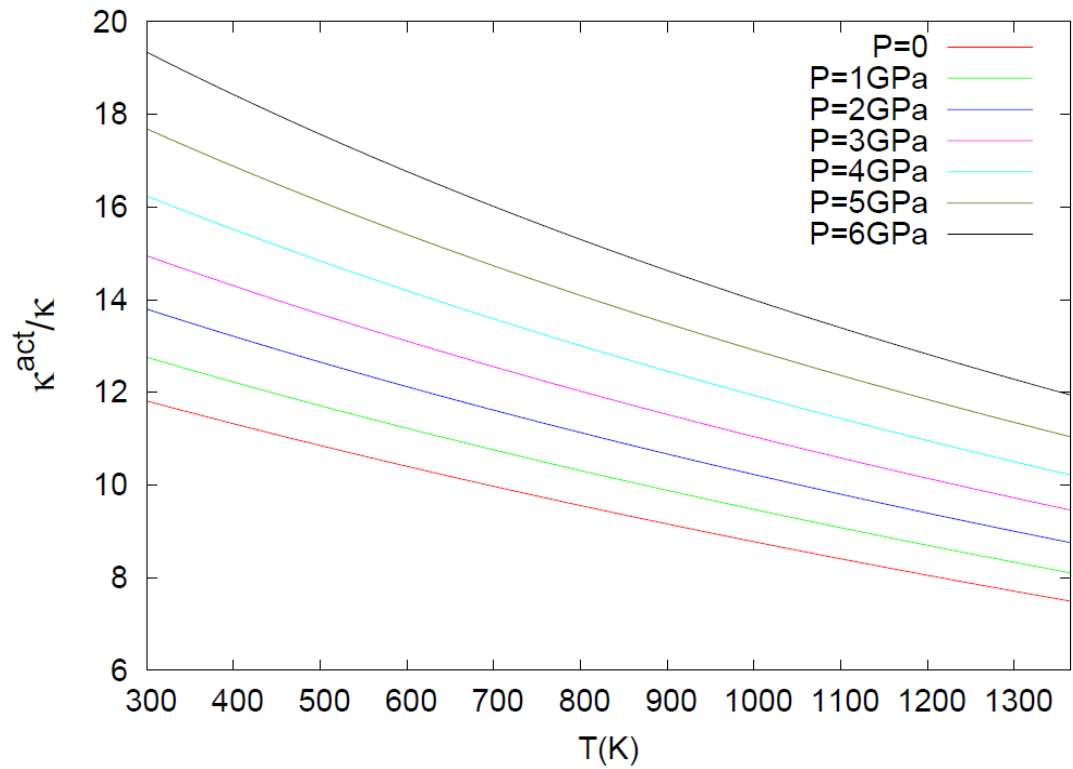

**Fig.1.** The values of the ratio $\dfrac{\kappa^{act}}{\kappa}$ versus temperature for constant pressure. The results are plotted for several values of the pressure depicted with various colours (see the inset).

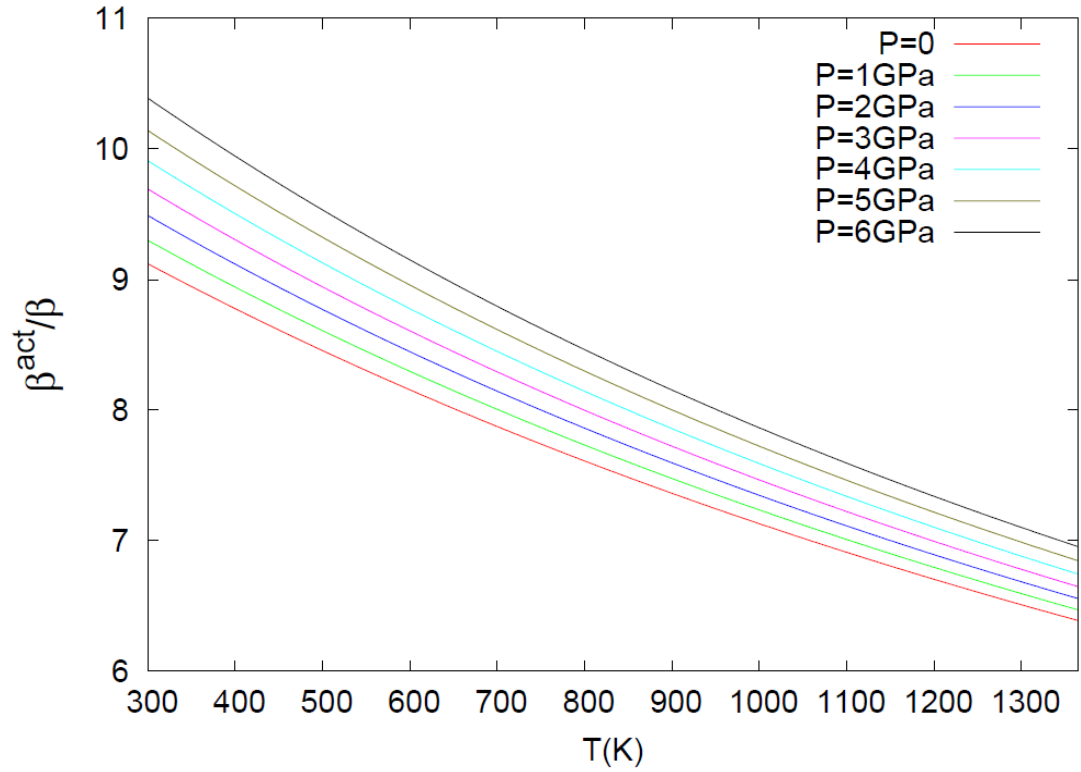

**Fig. 2** The values of the ratio $\dfrac{\beta^{act}}{\beta}$ versus temperature for constant pressure. The results are plotted for several values of the pressure depicted with various colours (see the inset).